\documentclass[%
 aip,
 amsmath,amssymb,
 reprint,%
]{revtex4-1}

\usepackage{graphicx}% Include figure files
\usepackage{dcolumn}% Align table columns on decimal point
\usepackage{bm}% bold math
\usepackage[utf8]{inputenc}
\usepackage[T1]{fontenc}
\usepackage{mathptmx}
\usepackage{etoolbox}

\makeatletter
\def\@email#1#2{%
 \endgroup
 \patchcmd{\titleblock@produce}
  {\frontmatter@RRAPformat}
  {\frontmatter@RRAPformat{\produce@RRAP{*#1\href{mailto:#2}{#2}}}\frontmatter@RRAPformat}
  {}{}
}%
\makeatother
\begin{document}

\preprint{AIP/123-QED}

\title[ ]{A Geometrical Design Tool for Building Cost-Effective Layout-Aware n-Bit Quantum Gates Using the Bloch Sphere Approach}
% Force line breaks with \\

\author{Ali Al-Bayaty}
 \affiliation{Department of Electrical and Computer Engineering, Portland State University, USA.}
 \email{albayaty@pdx.edu}
 
% \author{Shanyan Chen}
%  \affiliation{College of Information Engineering, Capital Normal University, China.}

% \author{Ali Al-Shuwaili}
%  \affiliation{College of Engineering, University of Information Technology \& Communications, Iraq.}

% \author{Abbas AlZubaidi}
%  \affiliation{College of Healthcare Technologies, American University of Iraq Baghdad, Iraq.}

\author{Marek Perkowski}
 \affiliation{Department of Electrical and Computer Engineering, Portland State University, USA.}

\date{30 December 2025}% It is always \today, today,

\begin{abstract}
The conventional design technique of any $n$-bit quantum gate is mainly achieved using unitary matrices multiplication, where $n \geq 2$ and $1 \leq m \leq n-1$ for $m$ target qubits and $n-m$ control qubits. These matrices represent quantum rotations by an $n$-bit quantum gate. For a quantum designer, such a conventional technique requires extensive computational time and effort, which may generate an $n$-bit quantum gate with a too high quantum cost. The Bloch sphere is only utilized as a visualization tool to verify the conventional design correctness for quantum rotations by a quantum gate. In contrast, this paper introduces a new concept of using the Bloch sphere as a ``geometrical design tool'' to build cost-effective $n$-bit quantum gates with lower quantum costs. This concept is termed the ``Bloch sphere approach (BSA)''. In BSA, a cost-effective $n$-bit quantum gate is built without using any unitary matrices multiplication. Instead, the quantum rotations for such a gate are visually selected using the geometrical planar intersections of the Bloch sphere. The BSA can efficiently map $m$ targets among $n-m$ controls for an $n$-bit quantum gate, to satisfy the limited layout connectivity for the physical neighboring qubits of a quantum computer. Experimentally, $n$-bit quantum gates built using the BSA always have lower quantum costs than those for such gates built using the conventional quantum design techniques.
\end{abstract}

\maketitle

% \begin{quotation}
% The ``lead paragraph'' is encapsulated with the \LaTeX\ 
% \verb+quotation+ environment and is formatted as a single paragraph before the first section heading. 
% (The \verb+quotation+ environment reverts to its usual meaning after the first sectioning command.) 
% Note that numbered references are allowed in the lead paragraph.
% %
% The lead paragraph will only be found in an article being prepared for the journal \textit{Chaos}.
% \end{quotation}

\section{Introduction}
In quantum computing, an $n$-bit quantum gate of $m$ target qubits and $n-m$ control qubits can be conventionally designed through the multiplication approach of its complete set of unitary matrices \cite{barenco,ralph,nielsen,lapierre}, where $n \geq 2$ and $1 \leq m \leq n-1$. Every unitary matrix represents a single quantum rotation for such an $n$-bit quantum gate. In general, some $n$-bit quantum gates are fundamentally used to construct other $n$-bit quantum gates. For instance, the $n$-bit Toffoli gate \cite{barenco,ralph,nielsen,lapierre} is mainly used to construct the $n$-bit Fredkin \cite{smolin1996five,patel2016quantum,gala,cala}, $n$-bit Miller \cite{gala,cala,lee2006cost,hung2006optimal}, and $n$-bit Peres and $n$-bit inverse Peres \cite{gala,cala,hung2006optimal,szyprowski2013low} gates. Subsequently, an $n$-bit quantum gate can be completely redesigned using the phase polynomial approach \cite{shende2008cnot,schmitt2022tweedledum}. Besides, the correctness of this phase-based approach can also be algebraically verified using unitary matrices multiplication.

Theoretically, on the one hand, Barenco et al. \cite{barenco} discussed the construction of the 3-bit Toffoli gate using the unitary matrices multiplication. On the other hand, Shende and Markov \cite{shende2008cnot} also constructed the 3-bit Toffoli gate using the phase polynomial approach. Both approaches aimed to design the quantum counterpart of a classical Boolean AND gate of two inputs, i.e., two control qubits, and one output, i.e., one target qubit. In addition, the correctness of these two conventional approaches can be visually verified using the geometrical quantum rotations on the Bloch sphere. Hence, the Bloch sphere is mainly utilized in quantum computing as the ``geometrical visualization tool'' for only verifying the final conventional design of any $n$-bit quantum gate.

Practically, we noticed that these two conventional approaches do not generate efficient $n$-bit quantum gates in the context of certain defined higher quantum costs. In this paper, the quantum cost is the total number of generated native (basis) gates for a transpiled (mapped) quantum circuit into the layout (architecture) of a real quantum computer, i.e., a quantum processing unit (QPU). The following three observations lead to proposing this research for minimizing quantum costs:

\begin{enumerate}
    \item Every QPU has a limited connectivity layout for a number of physical neighboring qubits, e.g., the \texttt{ibm\_torino} QPU of 133 qubits has the heavy-hex layout of up to four connected physical neighboring qubits \cite{baglio2024data,hung2025improved}. Notice that such a limited connectivity layout is not taken into account during the design stage, i.e., the conventional design for an $n$-bit quantum gate is not a layout-aware approach.
    \item Every QPU has a defined supported set of native gates that are generated in the final transpiled quantum circuits, e.g., the \texttt{ibm\_torino} QPU has the native gates of identity (I), Pauli-X (X), rotational X (RX), half-rotational X ($\sqrt{\text{X}}$), rotational Pauli-Z (RZ), rotational ZZ (RZZ), and controlled-Z (CZ), where their quantum rotations are defined in $\frac{\pi}{k}$ radians and $k \in \mathbb{R}$. Notice that, for current IBM QPUs, the I, X, RX, $\sqrt{\text{X}}$, and RZ are single-qubit native gates, and the RZZ and CZ are two-qubit native gates.
    \item An $n$-bit quantum gate is, generally, transpiled into a QPU with a large set of native gates due to the non-native gates decomposition. That means the final transpiled quantum circuit always has a higher quantum cost.
\end{enumerate}

Please observe that the limited connectivity layout of a QPU could add additional SWAP gates \cite{wille2014optimal,gokhale2024faster} to connect the physical non-neighboring qubits for a transpiled quantum circuit, which increases its quantum cost as well. A higher quantum cost could decrease the fidelity of a transpiled quantum circuit, by increasing the physical qubits decoherence that may introduce errors in the final measured results \cite{mi2022securing,murali2019noise,koch2020demonstrating}.

Based on the aforementioned three observations, the goal of this paper is to introduce a new geometrical design approach for building cost-effective $n$-bit quantum gates in the context of lower quantum costs and layout-aware efficient mapping into a specific QPU. This new geometrical design approach does not require any unitary matrices multiplication and phase polynomial generation techniques. Instead, our approach mainly utilizes the Bloch sphere itself as a ``geometrical design tool'' for visually building cost-effective $n$-bit quantum gates. This new approach is termed the ``Bloch sphere approach (BSA)''. Such that, in the BSA, the quantum rotations for the to-be-built $n$-bit quantum gate are visually selected using the geometrical planar intersections with the Bloch sphere. These geometrical planar intersections are the functional points in the XY-plane (equator) of the Bloch sphere. In our research, we classify these functional points into ``octants'', ``quadrants'', and ``semicircles'', due to the fact that all quantum rotations of the supported native gates are performed on this XY-plane for a specific QPU with the defined set of native gates and the layout geometry. For instance, the following native gates of the \texttt{ibm\_torino} QPU: (i) RZ($+\frac{\pi}{4}$) as `T' and RZ($-\frac{\pi}{4}$) as `$\text{T}^\dagger$' gates perform their quantum rotations on the octants, (ii) RZ($+\frac{\pi}{2}$) as `S' and RZ($-\frac{\pi}{2}$) as `$\text{S}^\dagger$' gates perform their quantum rotations on the quadrants, and (iii) RZ($+\pi$) as `Z' gates perform their quantum rotations on the semicircles of the XY-plane.

With respect to the layout geometry, the BSA effectively enhances the aforementioned three observations as follows.

\begin{enumerate}
    \item Mapping $m$ target qubits in the middle of $n-m$ control qubits of an $n$-bit quantum gate without adding SWAP gates to connect the physical non-neighboring qubits, to overcome the limited connectivity layout for a specific QPU.
    \item The generated $n$-bit quantum gate is initially built from the supported native gates of a specific QPU.
    \item The generated $n$-bit quantum gate is transpiled into a specific QPU with the same initial number of supported native gates, since no further gates decomposition is required. That means its final transpiled quantum circuit always has a lower quantum cost.
\end{enumerate}

Subsequently, using the BSA, various cost-effective $n$-bit quantum gates and libraries are designed for IBM quantum computers using the symmetrical circuits structure \cite{barenco,al2024cost,gala,cala,al2024p}, Clifford+T gates \cite{nielsen,cala,al2024p,bravyi2005universal}, and IBM native gates. Our two designed cost-effective $n$-bit quantum libraries are the generic architecture of layout-aware $n$-bit (GALA-$n$) operators \cite{gala} and Clifford+T-based architecture of layout-aware $n$-bit (CALA-$n$) operators \cite{cala}, which both consist of: (i) $n$-bit Toffoli gates, (ii) $n$-bit Boolean (AND, NAND, OR, NOR, Implication, and Inhibition) gates \cite{gala,cala,wakerly}, (iii) $n$-bit controlled-$\sqrt{\text{X}}$ (CV) and $n$-bit controlled-$\sqrt{\text{X}}^\dagger$ ($\text{CV}^\dagger$) gates \cite{barenco,gala,cala}, (iv) $n$-bit Fredkin gates, (v) $n$-bit Miller gates, and (vi) Boolean-Phase SWAP ``$p$-SWAP'' gate \cite{al2024p}, where $2 \leq n \leq 5$ qubits and $- \pi \leq p \leq + \pi$ radians. Notably, both GALA-$n$ and CALA-$n$ quantum libraries have become part of the IBM Qiskit Ecosystem \cite{ecosystem}.

In this research, we also introduce a new generic metric for efficiently calculating the quantum cost for the final transpiled quantum circuits into a specific QPU. This is an addition to several costs introduced in the past \cite{lee2006cost,maslov2003improved}. We term our metric the ``weighted transpilation quantum cost (WTQC)'', which is an improved variant of our previously proposed TQC in \cite{gala,al2025layout}. The WTQC is the weighted sum of four components: (i) the single-qubit native gates, (ii) the two-qubit native gates, (iii) the SWAP gates, and (iv) the depth for the final transpiled quantum circuit, where each component has its own weight ``$W_i$'' for importance ($W_i \geq 0$). Notably, the WTQC is a technology-dependent cost metric, while Maslov cost \cite{maslov2003improved} is a technology-independent cost metric that does not calculate the decomposed non-native gates after transpilation, e.g., the decomposed 3-bit Toffoli gates. Notice that Maslov cost is good for fast approximate cost evaluation, while the WTQC is more appropriate for evaluating a transpiled circuit for a specific QPU. Experimentally, after transpilation with an IBM QPU, we prove that different $n$-bit quantum gates and operators built using the BSA always have a lower WTQC than those gates built using the conventional design approaches.

\section{Preliminaries}

\subsection{The Bloch Sphere}
\begin{figure}
	\includegraphics[width=0.5\textwidth]{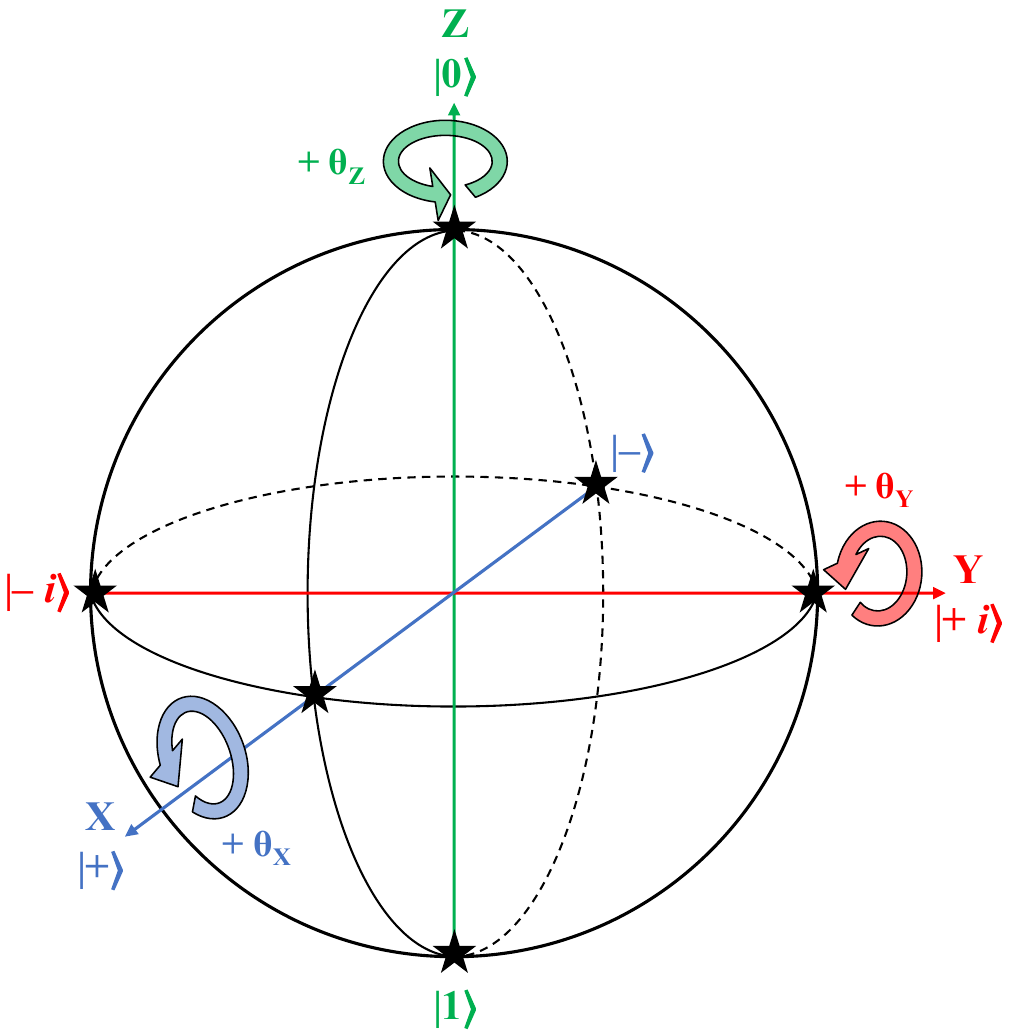}
	\caption{\label{fig:one}The Bloch sphere is a three-dimensional space with three spatial axes (X, Y, and Z) for illustrating six distinct base states (the stars) of a qubit: (i) the $|0\rangle$ and $|1\rangle$ states intersecting the Z-axis with the Bloch sphere, (ii) the $|+\rangle$ and $|-\rangle$ states intersecting the X-axis with the Bloch sphere, and (iii) the $|+i\rangle$ and $|-i\rangle$ states intersecting the Y-axis with the Bloch sphere.}
\end{figure}

The Bloch sphere is a geometrical sphere in a three-dimensional space with three X, Y, and Z spatial axes, which represents the quantum states of a qubit. When a series of quantum gates is applied to a qubit, the Bloch sphere visualizes such quantum operations in Hilbert space \cite{nielsen,lapierre}. Therefore, the Bloch sphere is commonly utilized in quantum computing as a geometrical visualization (and verification) tool. Fig. \ref{fig:one} depicts six distinct base states of a qubit, and each visualizes an intersection of one of the three spatial axes with the Bloch sphere.

An $n$-bit quantum gate having X (as a Pauli-X), Y (as a Pauli-Y), or Z (as a Pauli-Z) in its notation rotates the states of a qubit around the X, Y, or Z axis within the Bloch sphere, respectively, where $n \geq 1$. Such a quantum rotation around a spatial axis relies on a defined rotational angle ($\pm \theta$) expressed in radians, as shown in Fig. \ref{fig:one}, where $+ \theta$ represents a counterclockwise rotation and $- \theta$ represents a clockwise rotation for $-\pi \leq \theta \leq +\pi$. For instance, the following Z gates rotate the states of a qubit by the Z-axis for the defined $\theta$ angles:

\begin{itemize}
    \item The Z gate performs the quantum rotation of $+ \pi$ radians.
    \item The $\sqrt{\text{Z}}$ gate performs the quantum rotation of $+ \frac{\pi}{2}$ radians. The $\sqrt{\text{Z}}$ gate is the so-called `S' gate \cite{nielsen,lapierre,barenco}.
    \item The $\sqrt[4]{\text{Z}}$ gate performs the quantum rotation of $+ \frac{\pi}{4}$ radians. The $\sqrt[4]{\text{Z}}$ gate is the so-called `T' gate \cite{nielsen,lapierre,schmitt2022tweedledum}.
\end{itemize}

Notice that all $n$-bit quantum gates are unitary gates \cite{nielsen,lapierre,barenco}, and a few of them are non-Hermitian gates \cite{nielsen,lapierre,barenco}, i.e., their quantum operations are not equal to their own inverses, as the aforementioned S and T gates. In addition, some rotational gates alter the global phase for certain states of a qubit. Thus, the choice of rotational gates can be a critical factor in circuit design. For instance, the Z gate is not the same as the IBM native RZ($+ \pi$) gate, due to the global phase difference of $-i$ between the two gates. Such that, RZ($+ \pi$) = $e^{-i\frac{\pi}{2}Z} = \left[ \begin{array}{cc}
e^{-i\frac{\pi}{2}} & 0 \\ 0 & e^{+i\frac{\pi}{2}} \end{array} \right] = -i~\text{Z}$

\subsection{Clifford+T Gates and Symmetrical Circuit Structures}
The Clifford+T gates \cite{nielsen,cala,al2024p,bravyi2005universal} are a set of (i) single-qubit gates, which are I, X, Y, Z, Hadamard (H), $\sqrt{\text{X}}$, $\sqrt{\text{X}}^\dagger$, S, $\text{S}^\dagger$, T, and $\text{T}^\dagger$, and (ii) two-qubit gates, which are controlled-X (CX, CNOT, or Feynman), controlled-Y (CY), CZ, and SWAP. The Clifford+T gates play a significant role in quantum computing, because they: (i) invert the phase of a Pauli gate, and (ii) construct a Pauli gate from other Pauli gates, as expressed in Eq. (\ref{eq:one}) below and stated in Table \ref{tab:one}, where C is a Clifford+T gate, $\text{P}_\sigma$ is a Pauli gate, $\text{P}_\Sigma$ is the resultant Pauli gate, the symbol ($\dagger$) is the conjugate transpose (adjoint) operator, and the symbol ($\cdot$) is the matrix multiplication operator \cite{barenco,nielsen,lapierre,cala,al2024p,bravyi2005universal}.

\begin{equation}
	\text{C} \cdot \text{P}_\sigma \cdot \text{C}^\dagger = \text{P}_\Sigma
	\label{eq:one}
\end{equation}

\begin{table}
\caption{\label{tab:one}Transformation examples of Clifford+T gates for inverting the phases of Pauli gates and constructing other Pauli gates.}
	\begin{ruledtabular}
	\begin{tabular}{ccccccc}
		C & $\cdot$ & $\text{P}_\sigma$ & $\cdot$ & $\text{C}^\dagger$ & = & $\text{P}_\Sigma$ \\
		\hline
		H & $\cdot$ & Z & $\cdot$ & H & = & X \\
        S & $\cdot$ & X & $\cdot$ & $\text{S}^\dagger$ & = & Y \\
        H & $\cdot$ & X & $\cdot$ & H & = & Z \\
        Z & $\cdot$ & X & $\cdot$ & Z & = & $-\text{X}$ \\
        Z & $\cdot$ & Y & $\cdot$ & Z & = & $-\text{Y}$ \\
        X & $\cdot$ & Z & $\cdot$ & X & = & $-\text{Z}$ \\
	\end{tabular}
	\end{ruledtabular}
\end{table}

For different quantum computers, the Clifford+T gates (CTG) are mostly supported as native gates. For instance, based on the native gates of \texttt{ibm\_torino} QPU, the Clifford+T gates are limited to the set \{X, Z, H, $\sqrt{\text{X}}$, S, $\text{S}^\dagger$, T, $\text{T}^\dagger$, CZ\}, which is denoted here by $\text{CTG}_0$. On the one hand, the gates of $\text{CTG}_0$ rotating around the X-axis of the Bloch sphere are the X and $\sqrt{\text{X}}$ gates. On the other hand, the gates of $\text{CTG}_0$ rotating around the Z-axis of the Bloch sphere are the Z, S, $\text{S}^\dagger$, T, $\text{T}^\dagger$, and CZ gates. The H gate of $\text{CTG}_0$ is primarily used in quantum computing to create superposition states, by transforming the state of a qubit from the Z-axis ($|0\rangle$ or $|1\rangle$) to the XY-plane ($|+\rangle$ or $|-\rangle$), respectively, and vice versa. In general, the H, $\sqrt{\text{X}}$, and $\sqrt{\text{X}}^\dagger$ can be utilized as superposition gates. However, different analyses should be taken into account when utilizing $\sqrt{\text{X}}$ and $\sqrt{\text{X}}^\dagger$ as superposition gates to build cost-effective $n$-bit gates for various quantum computers \cite{al2023superposition}.

Notably, for current IBM QPUs, the CNOT gate is a two-qubit non-native gate, this gate can then be decomposed into the set of Clifford gates \{H, CZ, H\}, where the H gates are placed for the target qubit. The H gate is also a single-qubit non-native gate that can be decomposed into the set of Clifford gates \{S, $\sqrt{\text{X}}$, S\}. All these decomposed Clifford gates are IBM native gates.

For $n \geq 2$ qubits, an $n$-bit quantum gate has $n-1$ controls (input) qubits and one target (output) qubit, except for the SWAP, Fredkin (CSWAP), and developer's specific-purpose gates. All $n-1$ controls are connected to one target. For symmetrical circuit structures, by routing (re-mapping) the target among $n-1$ controls, an $n$-bit quantum gate can then be cost-effectively fitted into the limited connectivity layout of the physical neighboring qubits for any quantum computer. Therefore, for such an $n$-bit quantum gate, SWAP gates are never utilized to connect the physical non-neighboring qubits of a quantum computer, and the final quantum cost for its transpiled quantum circuit is significantly decreased.

\begin{figure*}
    \centering
    \begin{minipage}[c]{0.4\textwidth}
        \centering
		\includegraphics[scale=0.5]{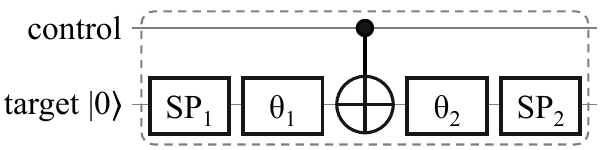}\\
        (a)
	\end{minipage}
    %\hfill
    \begin{minipage}[c]{0.5\textwidth}
        \centering
		\includegraphics[scale=0.5]{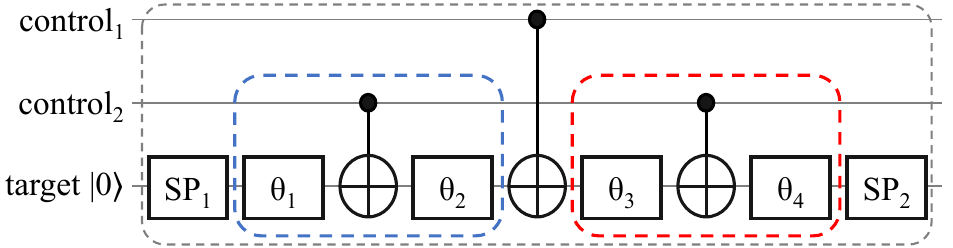}\\
        (b)
	\end{minipage}

    \begin{minipage}[c]{1\linewidth}
        \centering
		\includegraphics[scale=0.5]{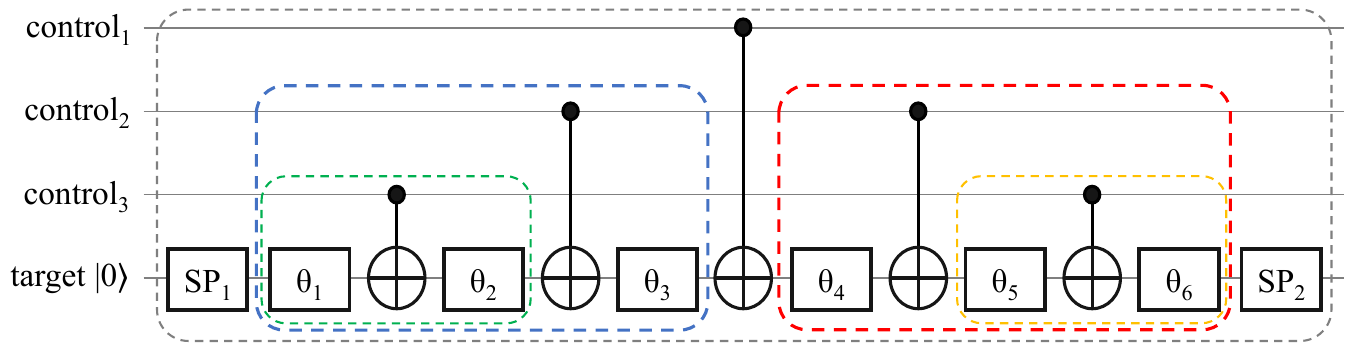}\\
        (c)
	\end{minipage}

    \begin{minipage}[c]{1\linewidth}
        \centering
		\includegraphics[scale=0.5]{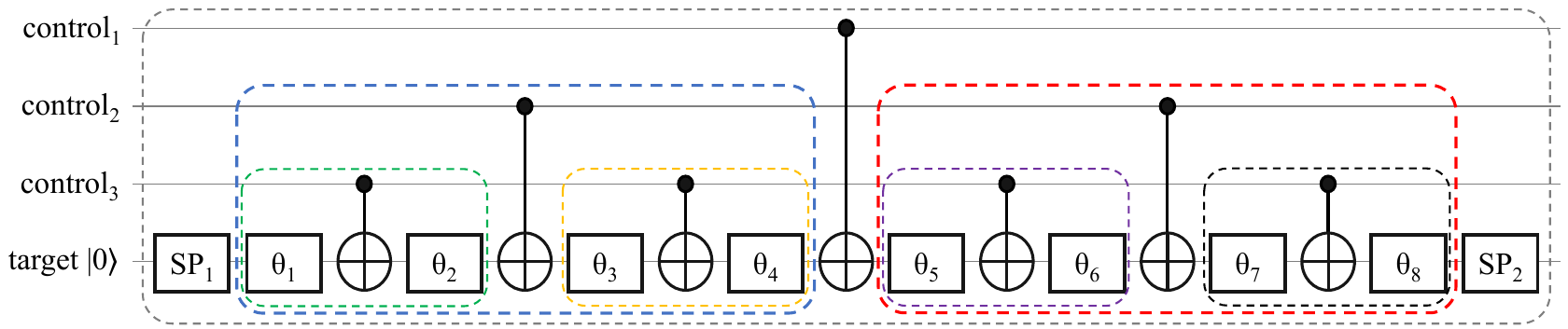}\\
        (d)
	\end{minipage}

	\caption{\label{fig:two}Arbitrary $n$-bit symmetrical quantum gates: (a) a 2-bit symmetrical gate, (b) a 3-bit symmetrical gate, (c) a 4-bit symmetrical gate, and (d) a 4-bit symmetrical gate, where $2 \leq n \leq 4$ and the colored rectangles denote symmetries \cite{al2025layout,al2024bsa}.}
\end{figure*}

Fig. \ref{fig:two} demonstrates arbitrary $n$-bit symmetrical quantum gates for $2 \leq n \leq 4$ qubits, where $\text{SP}_1$ and $\text{SP}_2$ are the superposition gates of H, $\sqrt{\text{X}}$, and $\sqrt{\text{X}}^\dagger$, $\theta$'s are the Clifford+T gates \{Z, S, $\text{S}^\dagger$, T, $\text{T}^\dagger$\} and the rotational angles of IBM native RZ($\pm \theta$) gates, and all $\theta$'s can have the same or different values.

\section{Methods}

\subsection{The Bloch Sphere Approach (BSA)}
\begin{figure*}
    \centering
    \begin{minipage}[c]{0.3\textwidth}
        \centering
		\includegraphics[scale=0.35]{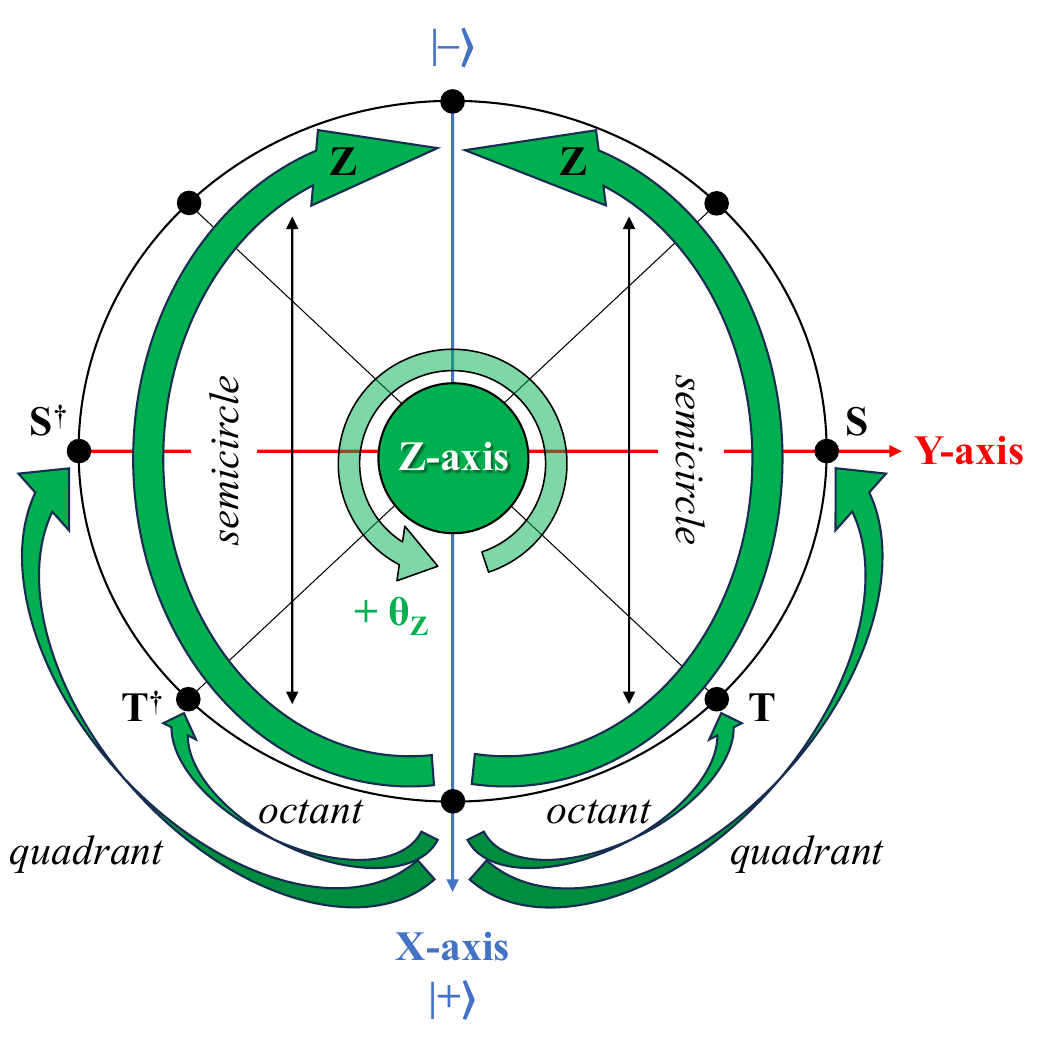}\\
        (a)
	\end{minipage}
    \hfill
    \begin{minipage}[c]{0.3\textwidth}
        \centering
		\includegraphics[scale=0.35]{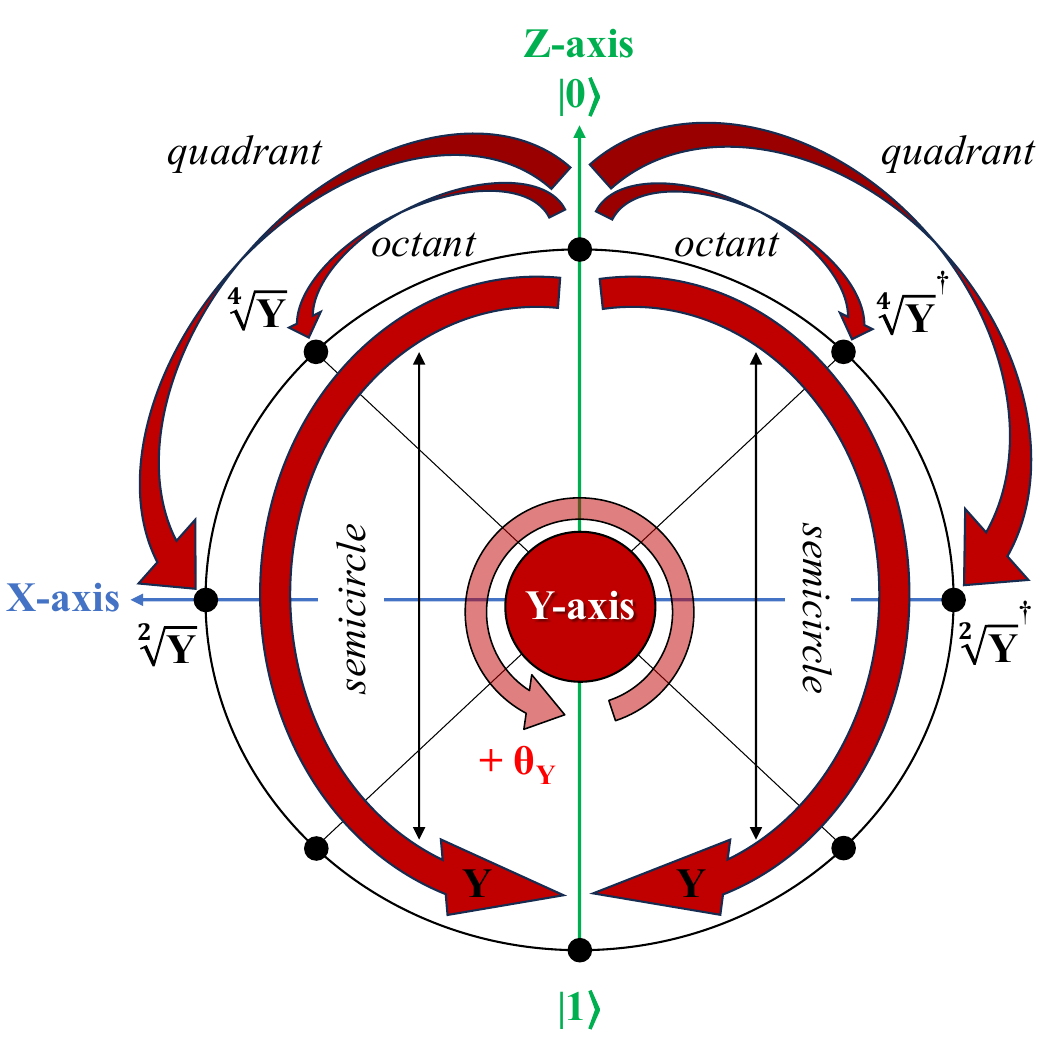}\\
        (b)
	\end{minipage}
    \hfill
    \begin{minipage}[c]{0.3\textwidth}
        \centering
		\includegraphics[scale=0.35]{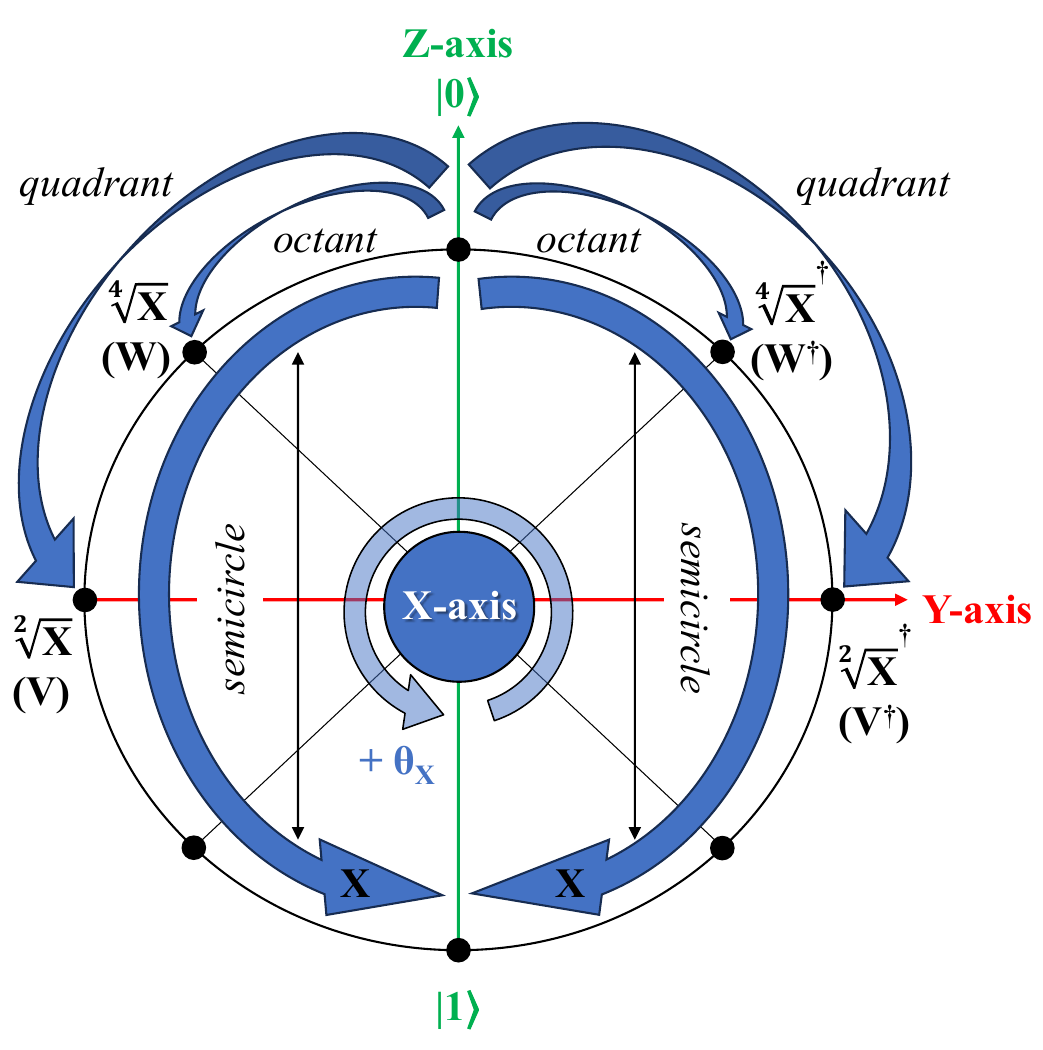}\\
        (c)
	\end{minipage}
    \vfill

    \begin{minipage}[c]{0.3\textwidth}
        \centering
		\includegraphics[scale=0.35]{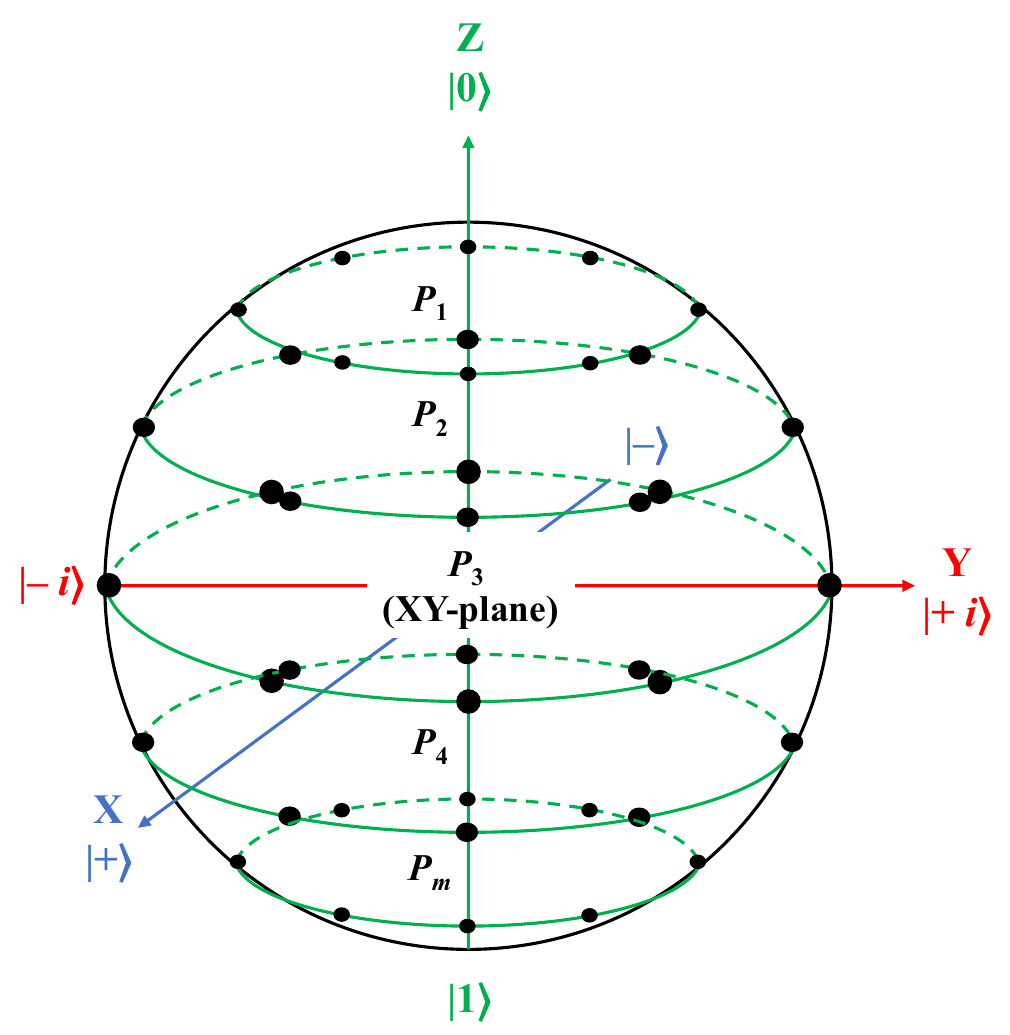}\\
        (d)
	\end{minipage}
    \hfill
    \begin{minipage}[c]{0.3\textwidth}
        \centering
		\includegraphics[scale=0.35]{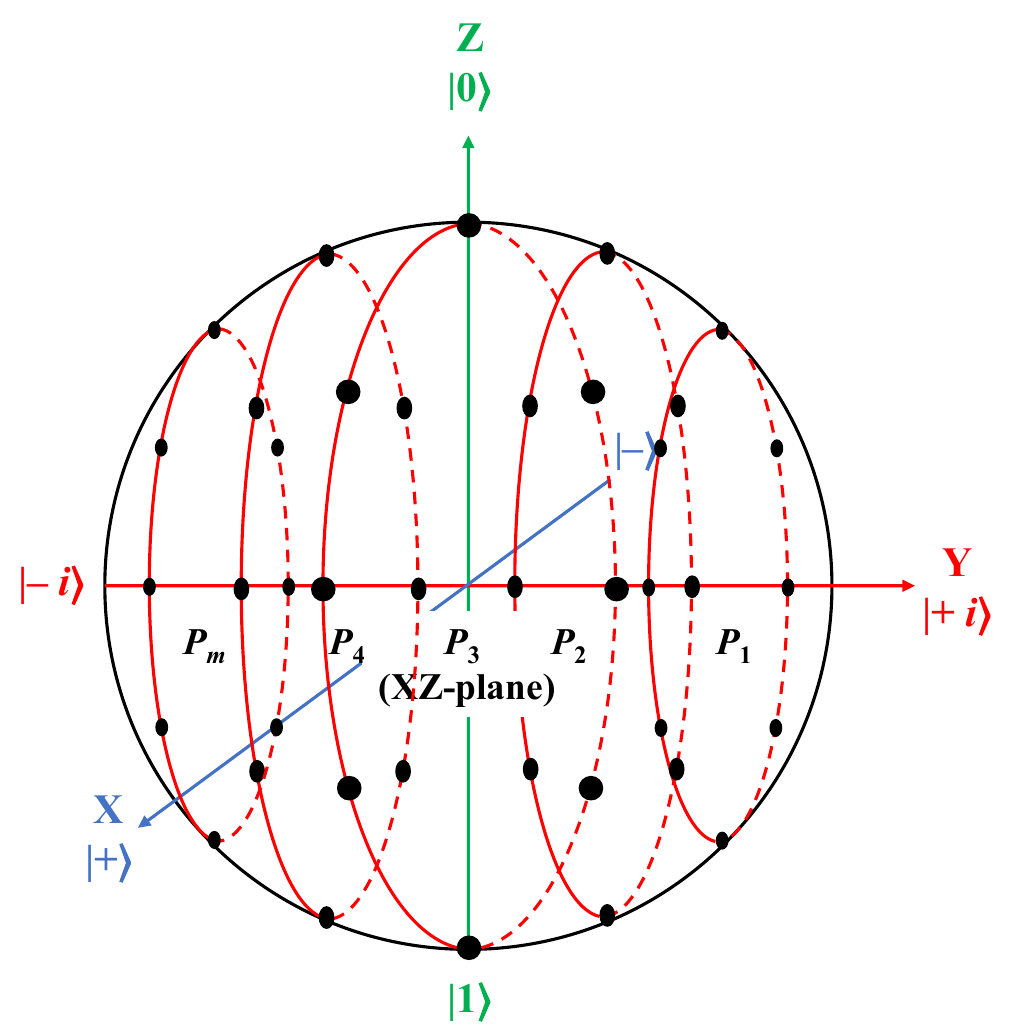}\\
        (e)
	\end{minipage}
    \hfill
    \begin{minipage}[c]{0.3\textwidth}
        \centering
		\includegraphics[scale=0.35]{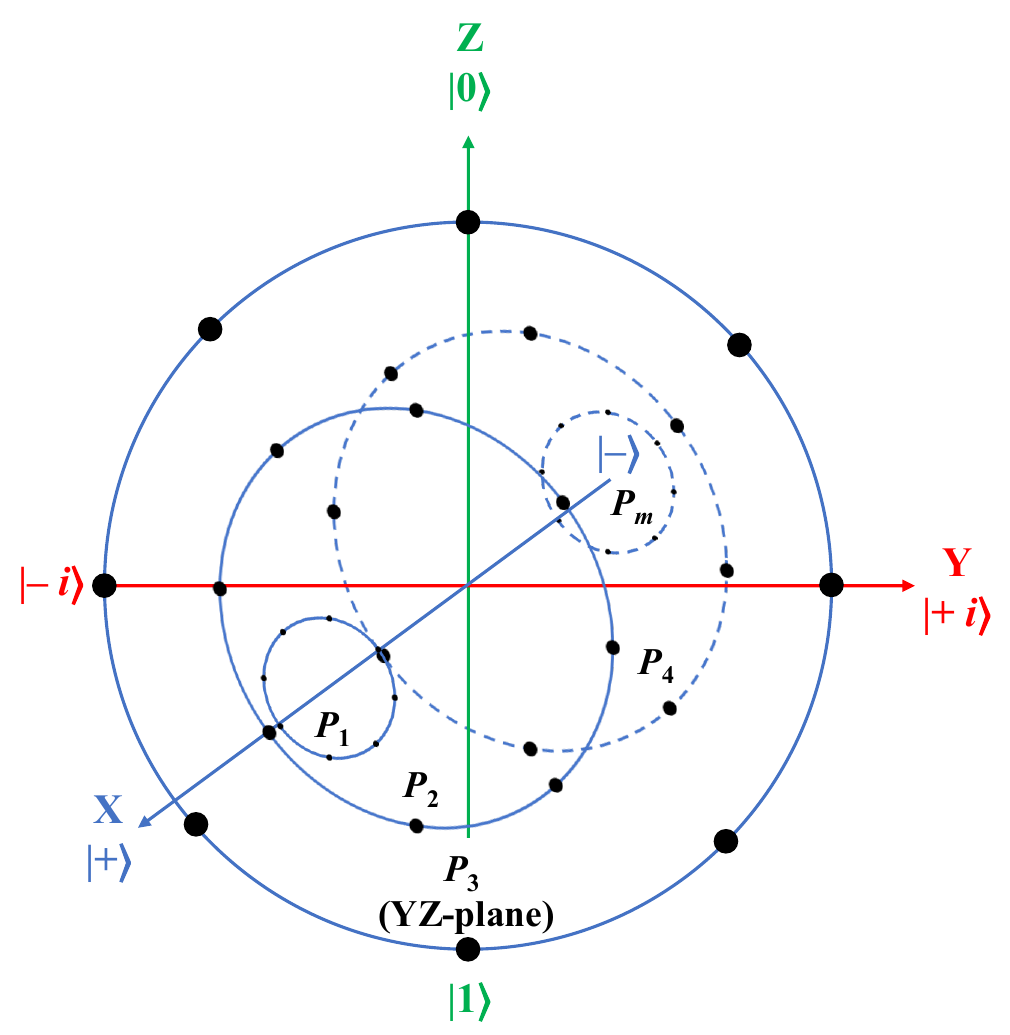}\\
        (f)
	\end{minipage}
    \vfill

    \begin{minipage}[c]{0.3\textwidth}
        \centering
		\includegraphics[scale=0.35]{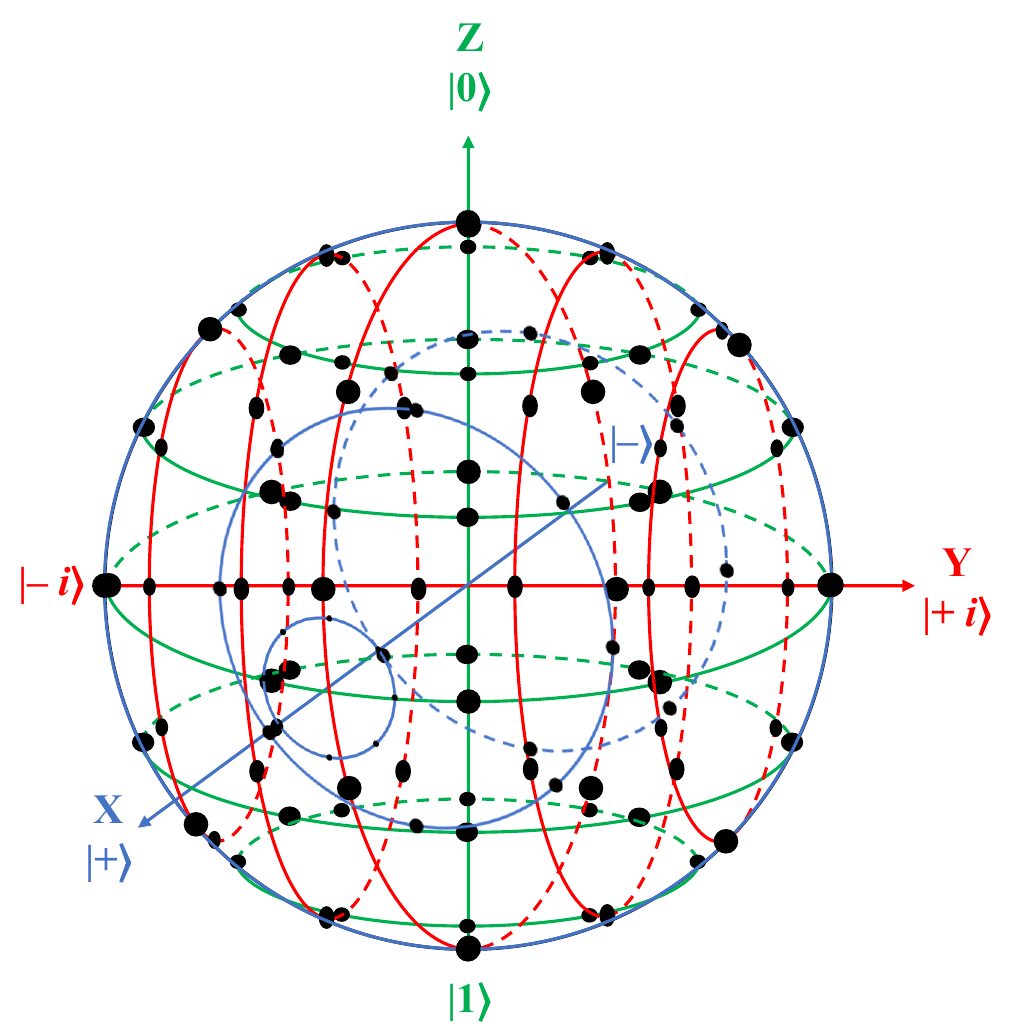}\\
        (g)
	\end{minipage}    

	\caption{\label{fig:three}Schematics of coordinate and parallel planes ($P$) perpendicular to various axes intersected with the Bloch sphere: (a) the XY-plane perpendicular to the Z-axis intersected with the Bloch sphere, as the $P_3$ in (d), (b) the XZ-plane perpendicular to the Y-axis intersected with the Bloch sphere, as the $P_3$ in (e), (c) the YZ-plane perpendicular to the X-axis intersected with the Bloch sphere, as the $P_3$ in (f), (d) $m$ parallel planes perpendicular to the Z-axis, (e) $m$ parallel planes perpendicular to the Y-axis, (f) $m$ perpendicular to the X-axis, and (g) three collections of $m$ parallel planes, each collection perpendicular to one of the three axes of the Bloch sphere, where $m \geq 1$ and the black dots subdividing their axial intersections with the Bloch sphere into the semicircles, quadrants, and octants \cite{al2024p,al2025layout,al2024bsa}.}
\end{figure*}

Geometrically, the coordinate and parallel planes are two-dimensional planes perpendicular to the individual three-coordinate spatial axes within the Bloch sphere. Fig. \ref{fig:three} illustrates different coordinate and parallel planes, as the XY-plane, XZ-plane, YZ-plane, and $m$ parallel planes ($P$) perpendicular to various of the three coordinate axes with the Bloch sphere, where $m \geq 1$ and the black dots denote the segments of ``semicircles'', ``quadrants'', and ``octants'' for the resulting circular intersections with the Bloch sphere.

In this paper, we introduce a new methodology using the intersections of different coordinate and parallel planes to geometrically design various cost-effective n-bit quantum gates. Hence, our methodology utilizes the Bloch sphere as a ``geometrical design tool'' to visually construct $n$-bit quantum gates. We termed this methodology the ``Bloch sphere approach (BSA)''. The cost-effectiveness of the BSA mainly relies on the aforementioned three observations discussed in the Introduction section, utilization of Clifford+T gates, and symmetrical circuit structures.

In \cite{gala,cala,al2024cost,al2024p,al2025layout,al2024bsa}, we discussed how to visually design any $n$-bit quantum gates using the BSA for IBM quantum computers. Because the quantum operations of all IBM native gates mainly rotate the states of a qubit around the X-axis and Z-axis of the Bloch sphere, we utilize the XY-plane of the Bloch sphere (shown in Fig. \ref{fig:three}(a)) to visually design different $n$-bit quantum gates. The intersections of the XY-plane with the Bloch sphere are divided into a number of segments to represent the quantum rotations ($\pm \theta$) of IBM native RZ gates applied to a qubit, as stated in three groups below, where $- \pi \leq \theta \leq + \pi$ radians. We choose a quantum base state on each circle of intersection and use the images of that state under various quantum rotations to subdivide the circle of intersection in various ways.

\begin{enumerate}
    \item The ``semicircle'' segment is half of the XY-plane's intersection with the Bloch sphere that represents the quantum rotations of Z gates, i.e., we use our base state’s images under the RZ($\pm \pi$) gate to subdivide our intersection circle. Here, we then consider our intersection circle to consist of two semicircles.
    \item The ``quadrant'' segments are obtained by replacing the RZ($\pm \pi$) gate in group (1) above, with either of the S and $\text{S}^\dagger$ gates, i.e., RZ($+ \frac{\pi}{2}$) and RZ($- \frac{\pi}{2}$) and their repetition of gates, respectively. Here, our intersection circle consists of four quadrants.
    \item The ``octant'' segments are obtained by replacing the RZ($\pm \pi$) gate in group (1) above, with either of the T and $\text{T}^\dagger$ gates, i.e., RZ($+ \frac{\pi}{4}$) and RZ($- \frac{\pi}{4}$) and their repetition of gates, respectively. Here, our intersection circle consists of eight octants.
\end{enumerate}

Notably, in the XY-plane, the IBM native X, $\sqrt{\text{X}}$, and RX gates rotate the states of a qubit around the X-axis by $+ \pi$,  $+ \frac{\pi}{2}$, and $\pm \theta$ radians, respectively. However, the IBM native CNOT gate rotates the state of its target qubit around the X-axis by $+ \pi$ radians, when its control qubit is set to the state of $|1\rangle$. Otherwise, no rotations occur.

Generally, the BSA can also be utilized to build generic and cost-effective $n$-bit quantum gates for other quantum computers, e.g., Intel, Google, and Rigetti, using other coordinate and parallel planes with the Bloch sphere, e.g., the XZ-plane and YZ-plane as demonstrated in Fig. \ref{fig:three}(b) and Fig. \ref{fig:three}(c), respectively, based on the supported Clifford+T gates as native gates for such quantum computers. Therefore, we introduce the BSA as a generic and open geometrical framework for prospective quantum computing research to build complex, interesting, and innovative $n$-bit quantum gates, circuits, and libraries.

\subsection{The BSA Design Steps}
In the BSA, the following steps express how to geometrically design cost-effective $n$-bit quantum gates using the XY-plane, Clifford+T gates (CTG), and symmetrical circuit structures (see Fig. \ref{fig:two}) for $n-1$ controls and one target, where $n \geq 2$ qubits.

\textbf{Step 1:} Describe the desirable operation (Boolean logic and quantum behavior) of the to-be-built $n$-bit gate using the truth table \cite{wakerly}, Karnaugh map \cite{wakerly}, binary decision diagram (BDD) \cite{ebendt2005advanced,wille2010effect}, just to name a few.

\textbf{Step 2:} Transform the quantum state of the target from the Z-axis of the Bloch sphere into the XY-plane using one H gate. Notice that the target is initially set to either $|0\rangle$ or $|1\rangle$ state, depending on the gate's operational purpose. The H gate can be decomposed into a sequence of three Clifford gates \{S, $\sqrt{\text{X}}$, S\}.

\textbf{Step 3:} For the target, define all Clifford+T gates as the initial set $\text{CTG}_0$ = \{H, X, $\sqrt{\text{X}}$, Z, S, $\text{S}^\dagger$, T, $\text{T}^\dagger$, CNOT\}. The CNOT gate can be generated in the sequence \{H, CZ, H\}.

\textbf{Step 4:} If the target is not controlling other qubits, then $\text{CTG}_0$ is limited to the set $\text{CTG}_1$ = \{H, X, $\sqrt{\text{X}}$, Z, S, $\text{S}^\dagger$, T, $\text{T}^\dagger$\}. Otherwise, $\text{CTG}_1$ = $\text{CTG}_0$.

\textbf{Step 5:} For the target, define all segments of the XY-plane as the initial set $\text{SEG}_0$ = \{semicircles, quadrants, octants\}.

\textbf{Step 6:} Since $\text{SEG}_0$ only identifies quantum gates rotating around the Z-axis of the Bloch sphere, then $\text{CTG}_1$ is limited to $\text{CTG}_2$ = \{Z, S, $\text{S}^\dagger$, T, $\text{T}^\dagger$\}. Assume that the quantum rotations around the X-axis of the Bloch sphere are not required for the target. Otherwise, this step is negligible and $\text{CTG}_2$ = $\text{CTG}_1$, but this will affect the selection of quantum rotations in the next steps for different quantum computers.

\textbf{Step 7:} Let $n_{CNOT}$ counts the total number of CNOT gates from the controls to the target, and the new sets $\text{CTG}_3$ and $\text{SEG}_1$ will be defined as follows.

\begin{itemize}
    \item If $n_{CNOT} = 1$, then $\text{CTG}_3$ = \{S, $\text{S}^\dagger$, T, $\text{T}^\dagger$\} and $\text{SEG}_1$ = \{quadrants, octants\}.
    \item If $n_{CNOT} = 2$, then a set of IBM native RZ gates defines $\text{CTG}_3$ = \{RZ($\theta$) $|~\theta \leq \pm \frac{\pi}{3}$\} for arbitrary $\text{SEG}_1$.
    \item If $n_{CNOT} = 3$, then $\text{CTG}_3$ = \{T, $\text{T}^\dagger$\} and $\text{SEG}_1$ = \{octants\}.
    \item If $n_{CNOT} > 3$, then a set of IBM native RZ gates defines $\text{CTG}_3$ = \{RZ($\theta$) $|~\theta \leq \pm \frac{\pi}{n_{CNOT}+1}$\} for arbitrary $\text{SEG}_1$.
\end{itemize}

\textbf{Step 8:} The $n_{CNOT}$ and the sequence of the permutative gates of $\text{CTG}_3$ should match the described quantum operation (logic and behavior) in \textbf{Step 1} above, which can be designed and visualized based on the found $\text{SEG}_1$ of the XY-plane.

\textbf{Step 9:} Re-transform the final reached quantum state of the target from the XY-plane into the Z-axis of the Bloch sphere using another H gate.

\textbf{Step 10:} Finally, the re-transformed quantum state of the target is considered the output of such a gate. For instance, the target in the state of $|1\rangle$ indicates a solution, while the state of $|0\rangle$ indicates a non-solution.

\subsection{Weighted Transpilation Quantum Cost and Layouts}
In \cite{gala,al2025layout}, we discussed a new metric of the final quantum cost calculation for a transpiled quantum circuit into the limited connectivity layout of any quantum computer, which was termed the ``transpilation quantum cost (TQC)''. However, in this paper, we introduce a generic variant metric of the TQC, which is termed the ``weighted transpilation quantum cost (WTQC)''. The WTQC, as expressed in Eq. (\ref{eq:two}), is the weighted sum of four components (as listed below) for a final transpiled quantum circuit, where $W_i$ is the weighted importance for each $N_1$, $N_2$, $XC$, and $D$ components and $W_i \geq 0$.

\begin{figure}
	\includegraphics[width=0.5\textwidth]{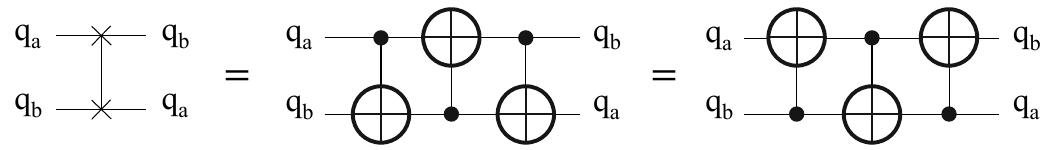}
	\caption{\label{fig:four}Schematics of a SWAP gate (left side) and its equivalent decompositions using three CNOT gates (middle and right sides), for swapping the states of two arbitrary indexed physical neighboring qubits ($\text{q}_\text{a}$ and $\text{q}_\text{b}$).}
\end{figure}

\begin{enumerate}
    \item $N_1$ is the total number of all single-qubit native gates in a transpiled circuit.
    \item $N_2$ is the total number of all two-qubit native gates in a transpiled circuit.
    \item $XC$ is the total number of all decomposed SWAP gates (see Fig. \ref{fig:four}), as the crossing connections ($XC$) among the utilized physical neighboring and non-neighboring qubits.
    \item $D$ is the depth, as the critical longest path of $N_1$, $N_2$, and $XC$, of a transpiled circuit.
\end{enumerate}

\begin{equation}
	\text{WTQC} = W_1 \cdot N_1 + W_2 \cdot N_2 + W_3 \cdot XC + W_4 \cdot D
	\label{eq:two}
\end{equation}

\begin{figure}
	\includegraphics[width=0.5\textwidth]{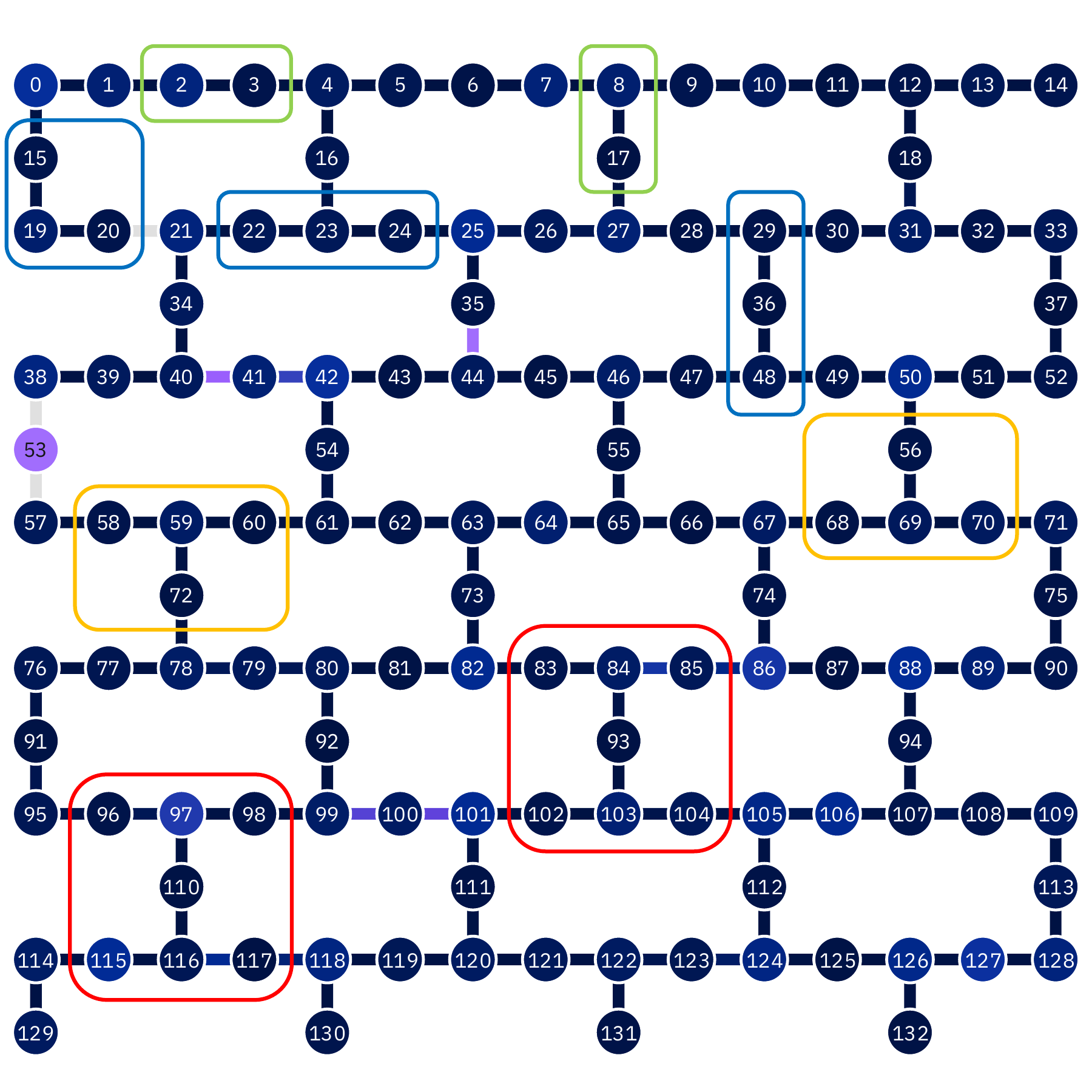}
	\caption{\label{fig:five}The heavy-hex layout of \texttt{ibm\_torino} QPU of 133 qubits \cite{baglio2024data,hung2025improved} illustrating arbitrary sets of $n$ physical neighboring qubits: (i) sets of two physical neighboring qubits ($n = 2$) as denoted in green, (ii) sets of three physical neighboring qubits ($n = 3$) as denoted in blue, (iii) sets of four physical neighboring qubits ($n = 4$) as denoted in orange, and (iv) sets of seven physical neighboring qubits ($n = 5$ with two additional ancilla qubits \cite{cala}) as denoted in red.}
\end{figure}

Notably, for all $W_i = 1$ in Eq. (\ref{eq:two}), if an $n$-bit quantum gate is designed using the BSA, the WTQC for its final transpiled quantum circuit can then be immediately predicted for a specific QPU, since such a gate was initially constructed using the supported native gates for that QPU. If an $n$-bit quantum gate only utilizes a specific set of physical neighboring qubits, then $XC = 0$ and $W_4$ is negligible. For instance, Fig. \ref{fig:five} illustrates arbitrary sets of $n$ physical neighboring qubits for the heavy-hex layout of \texttt{ibm\_torino} QPU of 133 qubits, where (i) $2 \leq n \leq 5$ qubits, (ii) the colors of physical qubits (circles) indicate the characteristics of single-qubit native gates, such as the T1 (relaxation time in $\mu\text{s}$), T2 (decoherence time in $\mu\text{s}$), readout errors, etc., (iii) the colors of channels (lines) indicate the characteristics of two-qubit native gates, such as gates length (in ns) and errors, and (iv) the circles and lines are scaled from lower (darker color) to higher (lighter color) values in points (ii) and (iii) above \cite{baglio2024data,hung2025improved}.

Based on Fig. \ref{fig:two}(b) and Fig. \ref{fig:five}, a symmetrical 3-bit quantum gate (two controls and one target) can be cost-effectively designed using the BSA, by geometrically mapping the two controls ($\text{control}_1$ and $\text{control}_2$) to the physical qubits indexed `15' and `20', respectively, and the target between these two controls to the physical qubit indexed `19', as denoted in the left-most blue group. Hence, the WTQC of this 3-bit gate has significantly decreased since $XC = 0$, i.e., SWAP gates are never added to connect these three physical neighboring qubits, with fewer generated $D$.

For the heavy-hex layout of any IBM QPU and $3 < n \leq 5$ qubits, symmetrical $n$-bit quantum gates can also be cost-effectively designed using the BSA, by selecting different sets of $n-1$ physical neighboring qubits for the controls and one physical neighboring qubit for the target that is mapped among these controls.

\section{Results and Discussion}
In our GALA-$n$ \cite{gala} and CALA-$n$ \cite{cala} quantum libraries, we discussed how to geometrically design cost-effective $n$-bit quantum gates and operators using the BSA, where $2 \leq n \leq 5$ qubits. For brevity and ease of demonstration, we present the aforementioned ten BSA design steps for geometrically constructing the cost-effective 3-bit Toffoli gate as follows.

\textbf{Step 1:} Describe the Boolean logic (as the quantum behavior) for the to-be-built 3-bit Toffoli gate using the truth table, as stated in Table \ref{tab:two}. Such a described gate only outputs the state of $|1\rangle$ when both controls are in the state of $|1\rangle$. Notice that because this gate only utilizes three qubits without any ancilla qubits, we can immediately design its quantum circuit shown in Fig. \ref{fig:two}(b), where its target is initially set to the state of $|0\rangle$.

\begin{table}
\caption{\label{tab:two}The quantum-based truth table for the Boolean logic of a 3-bit Toffoli gate of two control (input) qubits and one target (output) qubit.}
	\begin{ruledtabular}
	\begin{tabular}{cccc}
		$|\text{control}_2\rangle$ & $|\text{control}_1\rangle$ & $|\text{target}\rangle$ & Output in Boolean logic \\
		\hline
		$|0\rangle$ & $|0\rangle$ & $|0\rangle$ & False \\
        $|0\rangle$ & $|1\rangle$ & $|0\rangle$ & False \\
        $|1\rangle$ & $|0\rangle$ & $|0\rangle$ & False \\
        $|1\rangle$ & $|1\rangle$ & $|1\rangle$ & True \\
	\end{tabular}
	\end{ruledtabular}
\end{table}

\textbf{Step 2:} Transform the state of the target from the Z-axis of the Bloch sphere into the XY-plane using the H gate, as $\text{SP}_1$ shown in Fig. \ref{fig:two}(b).

\textbf{Step 3:} For the target, $\text{CTG}_0$ = \{H, X, $\sqrt{\text{X}}$, Z, S,$\text{S}^\dagger$, T, $\text{T}^\dagger$, CNOT\} for all $\theta$'s gates in Fig. \ref{fig:two}(b).

\textbf{Step 4:} Because the target is not controlling other qubits, $\text{CTG}_1$ = \{H, X, $\sqrt{\text{X}}$, Z, S, $\text{S}^\dagger$, T, $\text{T}^\dagger$\} for all $\theta$'s gates shown in Fig. \ref{fig:two}(b).

\textbf{Step 5:} For the target, $\text{SEG}_0$ = \{semicircles, quadrants, octants\}, as shown in Fig \ref{fig:three}(a).

\textbf{Step 6:} Step For the target, $\text{CTG}_2$ = \{Z, S, $\text{S}^\dagger$, T, $\text{T}^\dagger$\}, because there are no quantum rotation requirements by the X-axis of the Bloch sphere, for all $\theta$'s gates shown in Fig. \ref{fig:two}(b).

\textbf{Step 7:} $n_{CNOT} = 3$, then $\text{CTG}_3$ = \{T, $\text{T}^\dagger$\} for all $\theta$'s gates shown in Fig. \ref{fig:two}(b) and $\text{SEG}_1$ = \{octants\}.

\begin{figure*}
	\includegraphics[width=1.0\textwidth]{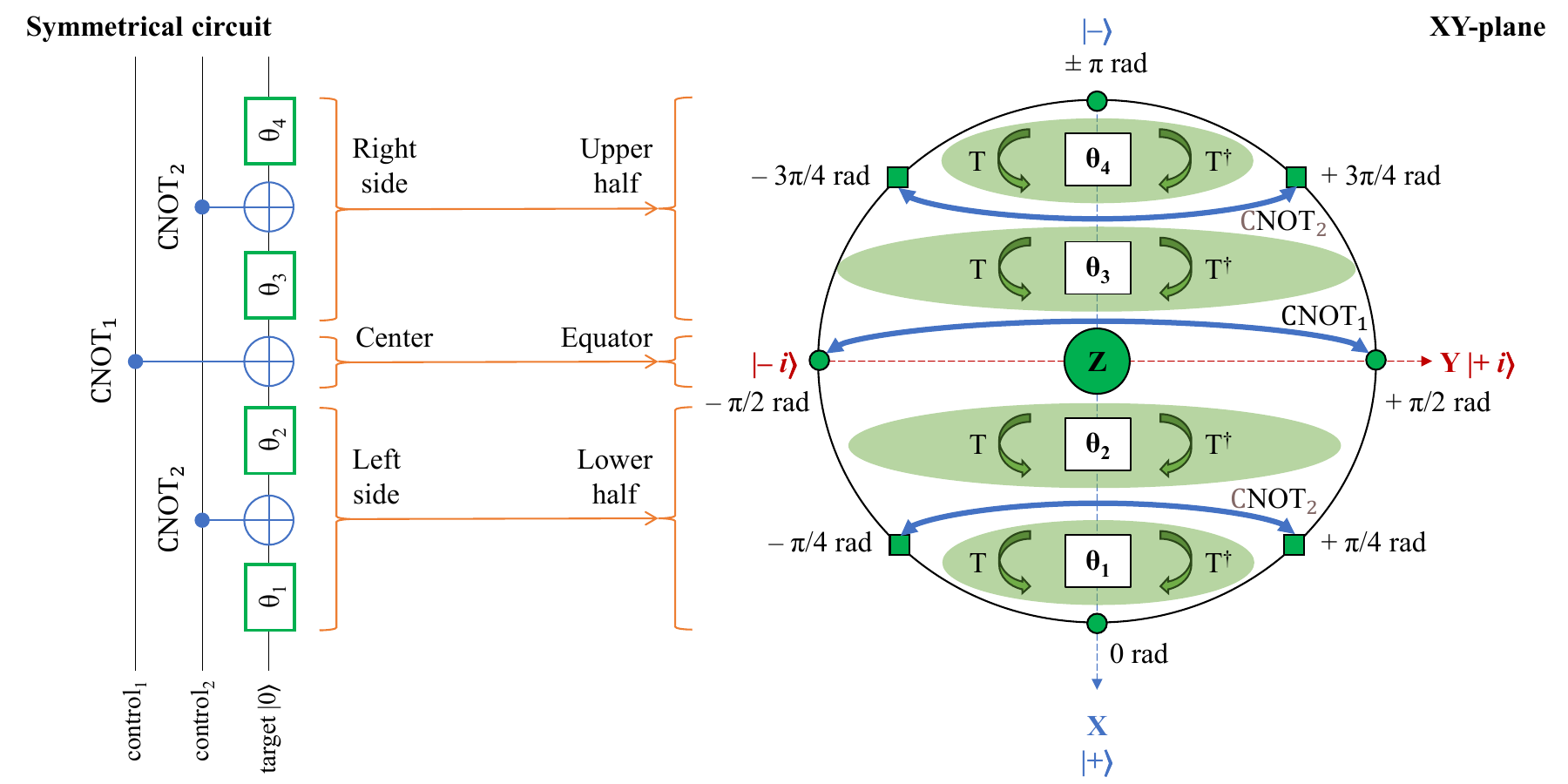}
	\caption{\label{fig:six}The geometrical matching visualization between the quantum gates of the symmetrical circuit of the 3-bit Toffoli gate and their quantum rotations in the XY-plane, where all quantum operations (gates and rotations) by the X-axis and the Z-axis are denoted in blue and green, respectively, and each $\theta$ has a set of permutative gates $\text{CTG}_3$ = \{T, $\text{T}^\dagger$\}. Notice that $\text{SP}_1$ and $\text{SP}_2$ (as H) gates are omitted in this figure, since the target is now in the XY-plane \cite{gala,cala,al2025layout}.}
\end{figure*}

\textbf{Step 8:} The $n_{CNOT}$ and the sequence of the permutative gates of $\text{CTG}_3$ should geometrically match the quantum-based Boolean logic expressed in Table \ref{tab:two}, and Fig. \ref{fig:six} geometrically visualizes this step.

\textbf{Step 9:} Based on Table \ref{tab:two} and Fig. \ref{fig:six}, the final reached state of the target is re-transformed from the XY-plane into the Z-axis of the Bloch sphere using the H gate, as the $\text{SP}_2$ shown in Fig. \ref{fig:two}(b).

\textbf{Step 10:} The re-transformed quantum state of the target is the output of this 3-bit Toffoli gate, i.e., the target in the state of $|1\rangle$ indicates a solution when both controls are in the state of $|1\rangle$. Otherwise, as a non-solution ($|0\rangle$) for all other combinations of both controls, as expressed in Table \ref{tab:two}.

Please observe that, in \textbf{Step 8} above, we found that the accurate permutative gates of $\text{CTG}_3$ for all $\theta$'s are in the sequence \{$\theta_1$ = $\text{T}^\dagger$, $\theta_2$ = T, $\theta_3$ = $\text{T}^\dagger$, $\theta_4$ = T\}, to successfully design this 3-bit Toffoli gate matching its quantum-based Boolean logic in Table \ref{tab:two}. Analytically, Table \ref{tab:three} verifies such accurate gate selections for $\text{CTG}_3$ for all $\theta$'s, where $\frac{\alpha~\pi}{\beta} = \frac{1}{\sqrt{2}} \left( |0\rangle + e^{~i \frac{\alpha~\pi}{\beta}} |1\rangle \right)$, $\alpha~\text{and}~\beta \in \mathbb{R}$, $e^{~\pm ix} = \text{cos(x)} \pm i~\text{sin(x)}$, and `–' means a gate is not applied to the target \cite{gala,cala,al2025layout}.

\begin{table*}
\caption{\label{tab:three}Analytical verifications for the accurate gate selections for all $\theta$'s of the 3-bit Toffoli gate \cite{gala,cala,al2025layout}.}
	\begin{ruledtabular}
	\begin{tabular}{ccccccccccc}
		$|\text{control}_2~\text{control}_1\rangle$ & $\text{SP}_1$ (H) & $\theta_1 (\text{T}^\dagger)$ & $\text{CNOT}_2$ & $\theta_2$ (T) & $\text{CNOT}_1$ & $\theta_3 (\text{T}^\dagger)$ & $\text{CNOT}_2$ & $\theta_4$ (T) & $\text{SP}_2$ (H) & Output in Boolean logic\\
		\hline
        $|0~0\rangle$ & $|+\rangle$ & $\frac{7\pi}{4}$ & – & $|+\rangle$ & – & $\frac{7\pi}{4}$ & – & $|+\rangle$ & $|0\rangle$ & False \\

        $|0~1\rangle$ & $|+\rangle$ & $\frac{7\pi}{4}$ & – & $|+\rangle$ & $|+\rangle$ & $\frac{7\pi}{4}$ & – & $|+\rangle$ & $|0\rangle$ & False \\

        $|1~0\rangle$ & $|+\rangle$ & $\frac{7\pi}{4}$ & $\frac{\pi}{4}$ & $|+ i\rangle$ & – & $\frac{\pi}{4}$ & $\frac{7\pi}{4}$ & $|+\rangle$ & $|0\rangle$ & False \\

        $|1~1\rangle$ & $|+\rangle$ & $\frac{7\pi}{4}$ & $\frac{\pi}{4}$ & $|+ i\rangle$ & $|- i\rangle$ & $\frac{5\pi}{4}$ & $\frac{3\pi}{4}$ & $|-\rangle$ & $|1\rangle$ & True \\
	\end{tabular}
	\end{ruledtabular}
\end{table*}

\begin{figure*}
	\includegraphics[width=1\textwidth]{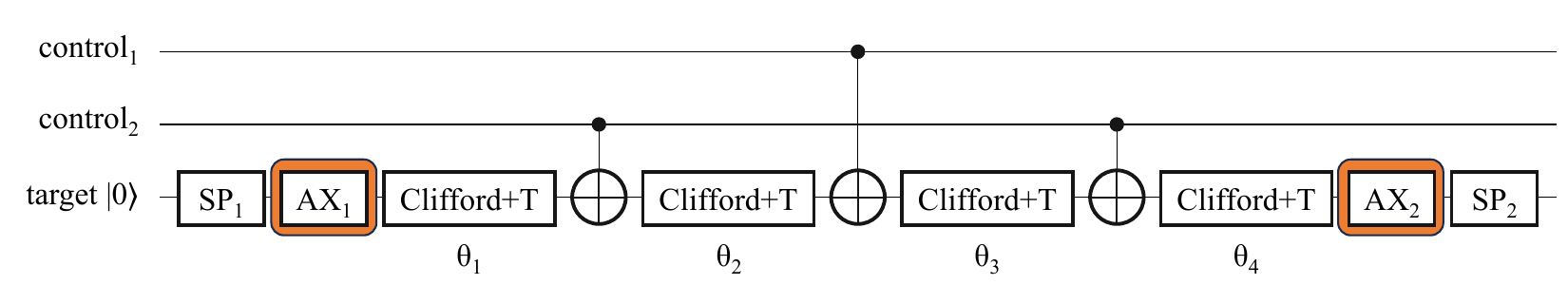}
	\caption{\label{fig:seven}Our generalized symmetrical circuit structure for the 3-bit quantum gate of CALA-$n$, where all $\theta$'s (as $\theta_1$, $\theta_2$, $\theta_3$, and $\theta_4$) are Clifford+T gates, $\text{SP}_1$ and $\text{SP}_2$ are the superposition gates, $\text{AX}_1$ and $\text{AX}_2$ are the auxiliary gates for constructing various 3-bit quantum Boolean-based gates, and the target qubit is initially set to the state of $|0\rangle$ \cite{cala,al2025layout}.}
\end{figure*}

In our research \cite{cala,al2025layout}, we also investigated other permutation orders and sequences for the gates selection of the found $\text{CTG}_3$ in \textbf{Step 8} above, by adding two auxiliary (AX) quantum gates to construct various 3-bit quantum Boolean-based gates as shown in Fig. \ref{fig:seven}. Subsequently, the constructed 3-bit quantum Boolean-based operators are the AND, NAND, OR, NOR, implication, and inhibition gates. Table \ref{tab:four} analytically states these permutation orders and sequences of $\text{CTG}_3$ for all $\theta$'s of a 3-bit symmetrical circuit shown in Fig. \ref{fig:seven}, where the target is initially set to the state of $|0\rangle$. Notice that, in Fig. \ref{fig:seven}, the $\text{AX}_1$ gate is reserved for future work and investigations to construct other useful quantum Boolean-based and phase-based operators.

\begin{table*}
\caption{\label{tab:four}The gate selections for constructing various 3-bit quantum Boolean-based operators using the BSA technique \cite{cala,al2025layout}.}
	\begin{ruledtabular}
	\begin{tabular}{rcccccccc}
		3-bit Boolean operators & $\text{SP}_1$ & $\text{AX}_1$ & $\theta_1$ & $\theta_2$ & $\theta_3$ & $\theta_4$ & $\text{AX}_2$ & $\text{SP}_2$  \\
		\hline
        AND: $(a \wedge b)$ & H & I & $\text{T}^\dagger$ & T & $\text{T}^\dagger$ & T & I & H \\

        NAND: $\neg~(a \wedge b)$ & H & I & $\text{T}^\dagger$ & T & $\text{T}^\dagger$ & T & $-$Z & H \\

        OR: $(a \vee b)$ & H & I & T & T & T & T & Z & H \\

        NOR: $\neg~(a \vee b)$ & H & I & T & T & T & T & I & H \\

        Implication: $(\neg~a \vee b)$ & H & I & $\text{T}^\dagger$ & $\text{T}^\dagger$ & T & T & $-$Z & H \\

        Inhibition: $\neg~(\neg~a \vee b)$ & H & I & $\text{T}^\dagger$ & $\text{T}^\dagger$ & T & T & I & H \\
	\end{tabular}
	\end{ruledtabular}
\end{table*}

Notably, in Table \ref{tab:four}, the quantum Boolean-based operators can be effectively used to build Boolean oracles \cite{figgatt2017complete,al2024concept,al2024grover} and Boolean-Hamiltonians transformed oracles \cite{al2024bht,al2024bht2}, especially for the functions and problems that are logically structured in the Product-of-Sums (POS) \cite{wakerly,huang2007note}, Sum-of-Products (SOP) \cite{wakerly,zimmermann2003optimized}, Exclusive-or Sum-of-Products (ESOP) \cite{mishchenko2001fast,sasao2002exmin2}, constraints satisfiable problems–satisfiability (CSP–SAT) \cite{perkowski2022inverse}, and in designing XOR functions of two product gates of different Hamming distances \cite{schaeffer2013synthesis}, just to name a few.

In this paper, the BSA along with the parametric selection and modification of the rotational angles ($\theta$'s), superposition gates ($\text{SP}_1$ and $\text{SP}_2$), and auxiliary gates ($\text{AX}_1$ and $\text{AX}_2$) are open topics for future quantum computing research, as an introductory geometrical design framework to construct interesting and cost-effective $n$-bit quantum Boolean-based and phase-based operators relying on the layouts and the number of $n$ physical neighboring qubits for different quantum computers.

Based on GALA-$n$ $(2 \leq n \leq 4)$ and CALA-$n$ $(2 \leq n \leq 5)$ quantum libraries, various experiments of $n$-bit quantum gates and operators were designed using conventional design approaches and the BSA. These experiments were then transpiled using the \texttt{ibm\_brisbane} QPU of 127 qubits \cite{farrell2024scalable}. The experiments built using conventional design approaches were transpiled using the standard IBM Transpiler \cite{wilson2020just}, i.e., the logical qubits of these experiments are automatically mapped to their corresponding physical qubits by this Transpiler. While the experiments built using the BSA were transpiled using our manual assignments for the logical qubits to their corresponding physical qubits based on the layout geometry of this QPU.

\begin{figure*}
	\includegraphics[width=1\textwidth]{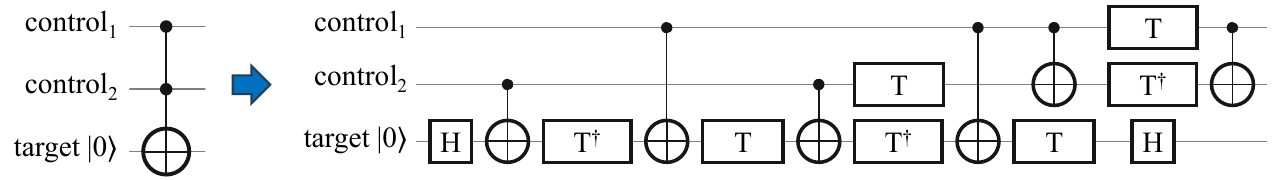}
	\caption{\label{fig:eight}The decomposition of an IBM non-native 3-bit Toffoli gate (left side) into a non-symmetrical circuit consisting of IBM native gates (right side). Notice that SWAP gates are required here to connect the controls and target (as the physical neighboring and non-neighboring qubits) to fit the layout of an IBM QPU (see Fig. \ref{fig:five}).}
\end{figure*}

Notice that the Transpiler mostly generates cost-expensive and non-symmetrical structures for transpiled circuits. For instance, Fig. \ref{fig:eight} illustrates the decomposition of the IBM non-native 3-bit Toffoli gate into a non-symmetrical circuit consisting of IBM native gates as part of IBM transpilation processes. For this reason, SWAP gates are normally added to connect the physical non-neighboring qubits (even for physical neighboring qubits) shown in Fig. \ref{fig:five}, yielding to increase the final quantum cost for such transpiled circuits. In contrast, we initially construct circuits using the symmetrical structures (see Fig. \ref{fig:two}) to be efficiently utilized by the BSA and to cost-effectively fit the layout geometry of a specific quantum computer. Additionally, our manual physical qubits assignment guarantees to never add SWAP gates. For this reason, the quantum costs of our transpiled circuits always have lower values than those generated by the Transpiler.

Table \ref{tab:five} summarizes the cost-effectiveness design approaches of the final transpiled circuits using our proposed WTQC metric, as expressed in Eq. (\ref{eq:two}), for the experiments generated from: (i) the conventional design approaches and transpiled using the Transpiler, and (ii) the BSA for symmetrical structures where their qubits are manually assigned and transpiled.

\begin{table*}
\caption{\label{tab:five}The WTQC of different transpiled circuits of $n$-bit quantum gates with the \texttt{ibm\_brisbane} QPU, where all weighted importance $W_i = 1$ and $2 \leq n \leq 5$ physical neighboring qubits.}
	\begin{ruledtabular}
	\begin{tabular}{cccccccccccc}
        % \multicolumn{No_of_columns}{c}{TEXT}
		Quantum& $n$ & \multicolumn{5}{c}{Cost of transpiled quantum circuits (conventional design approach)} & \multicolumn{5}{c}{Cost of transpiled quantum circuits (BSA)} \\
        operators & qubits & $N_1~+$ & $N_2~+$ & $XC~+$ & $D$ & = WTQC & $N_1~+$ & $N_2~+$ & $XC~+$ & $D$ & = WTQC \\ 
		\hline
        Controlled-$\sqrt{\text{X}}$ & 2 & 7 & 2 & 0 & 9 & 18 & 6 & 1 & 0 & 7 & 14 \\

        Controlled-$\sqrt{\text{X}}$ & 4 & 147 & 29 & 5 & 110 & 291 & 53 & 7 & 0 & 42 & 102 \\

        AND & 3 & 48 & 5 & 2 & 42 & 97 & 34 & 3 & 0 & 29 & 66 \\

        AND & 4 & 146 & 29 & 5 & 108 & 288 & 52 & 7 & 0 & 41 & 100 \\

        AND & 5 & 387 & 41 & 24 & 299 & 751 & 81 & 9 & 0 & 74 & 164 \\

        NAND & 4 & 105 & 20 & 2 & 83 & 210 & 52 & 7 & 0 & 41 & 100 \\

        OR & 4 & 102 & 20 & 2 & 80 & 204 & 52 & 7 & 0 & 41 & 100 \\

        OR & 5 & 398 & 41 & 27 & 304 & 770 & 92 & 9 & 0 & 76 & 177 \\

        NOR & 4 & 103 & 20 & 2 & 83 & 208 & 52 & 7 & 0 & 41 & 100 \\

        Implication & 3 & 49 & 9 & 1 & 39 & 98 & 28 & 3 & 0 & 21 & 52 \\

        Inhibition & 3 & 50 & 9 & 1 & 39 & 99 & 28 & 3 & 0 & 21 & 52 \\

        Fredkin & 3 & 67 & 7 & 4 & 59 & 137 & 35 & 5 & 0 & 22 & 62 \\

        Fredkin & 4 & 189 & 40 & 8 & 115 & 352 & 56 & 9 & 0 & 46 & 111 \\

        Miller & 4 & 151 & 29 & 3 & 113 & 296 & 70 & 13 & 0 & 58 & 141 \\		
	\end{tabular}
	\end{ruledtabular}
\end{table*}

Accordingly, our proposed BSA as the geometrical design tool successfully generates various cost-effective $n$-bit quantum gates using the symmetrical circuits structure, Clifford+T gates, as well as the supported native gates and layout geometry for a specific QPU. The BSA can provide a broad range of various cost-effective quantum Boolean-based gates and operators for practical applications in the fields of Boolean oracular and heuristic search algorithms, digital logic circuits, and machine learning, just to name a few. Our future work will concentrate on utilizing the BSA to cost-effectively re-develop our GALA-$n$ and CALA-$n$ quantum libraries to support various native gates and layouts of other superconducting quantum systems, such as Google and Rigetti.

\section{Conclusions}
This paper introduces a new quantum circuit design approach using the Bloch sphere as a ``geometrical design tool'' to build cost-effective $n$-bit quantum gates with lower quantum costs, where $n \geq 2$. This new design is termed the ``Bloch sphere approach (BSA)''. In the BSA, cost-effective $n$-bit quantum gates are built without using any unitary matrices multiplication, as this is a fundamental technique used in many conventional design approaches \cite{barenco,ralph,shende2008cnot,schmitt2022tweedledum} that may require extensive computational time and generate cost-expensive $n$-bit quantum gates. The BSA visually builds an $n$-bit quantum gate using the geometrical planar intersections (as the XY-plane) of the Bloch sphere, as the desirable quantum rotations for a to-be-built gate. To cost-effectively design $n$-bit quantum gates for $n-1$ control qubits and one target qubit, the BSA mainly utilizes these four factors: (i) the XY-plane, (ii) Clifford+T gates \cite{nielsen,cala,al2024p,bravyi2005universal}, (iii) symmetrical circuit structures \cite{barenco,al2024cost,gala,cala,al2024p}, as well as (iv) efficiently mapping the target qubit among the control qubits to geometrically satisfy the limited layout connectivity of $n$ physical neighboring qubits for a specific quantum computer.

In our research, we also introduce a new technology-dependent metric to calculate the quantum cost of the final transpiled $n$-bit quantum gates into a specific quantum computer. This new metric is termed the ``weighted transpilation quantum cost (WTQC)''. The WTQC is the weighted sum of: (i) the total number of single-qubit native gates, (ii) the total number of two-qubit native gates, (iii) the total number of SWAP gates among the utilized physical neighboring and non-neighboring qubits, and (iv) the depth of a transpiled quantum circuit. Experimentally, using an IBM quantum computer, various $n$-bit quantum gates designed using the BSA always have lower WTQC values than those for such gates designed using conventional design techniques.

In conclusion, on the one hand, the BSA as a geometrical design tool successfully creates different cost-effective $n$-bit quantum gates for IBM quantum computers, with the practical approval of using our introduced WTQC as a technology-dependent cost metric. On the other hand, the BSA along with different utilization of other geometrical planar intersections (as the XZ-plane and YZ-plane) of the Bloch sphere and other quantum rotational gates (including Clifford+T gates) is an open future quantum computing research in the topics of constructing many interesting and cost-effective $n$-bit quantum Boolean-based and phase-based operators for Boolean and phase oracles \cite{figgatt2017complete,al2024concept,al2024bht}, respectively, which are geometrical restricted to the limited layouts and the number of $n$ physical neighboring qubits for different quantum computers.

In our future research, we will fundamentally utilize the BSA to geometrically build cost-effective complex $n$-bit quantum operators, such as arithmetic adders, comparators, and symmetrical logical circuits \cite{chen2025stesso}. The WTQC can be formally used for the evaluation and benchmarking purposes of various cost-effective design approaches using different layouts of quantum computers. Based on our previous research, more design methods should be applied to other types of basic gates, such as the majority gate \cite{yang2005majority}, Peres and inverse Peres gates \cite{tsai2022methodologies}, and EXOR link (Hamming distance `HD') gates \cite{song2002minimization}, to develop and complete useful quantum libraries, especially for arithmetic circuits.

% \begin{acknowledgments}
% We wish to acknowledge the support of the author community in using
% REV\TeX{}, offering suggestions and encouragement, testing new versions,
% \dots.
% \end{acknowledgments}

\section*{Data Availability Statement}

The data that support the findings of this study are available within the article.

% References
\nocite{*}
\bibliography{aipsamp}% Produces the bibliography via BibTeX.

\end{document}